\documentclass[a4paper]{article}

\usepackage{xspace}
\usepackage[pdfpagelabels]{hyperref}
\usepackage{todonotes}

\usepackage{tikz}
\usepackage{pgfplots}
\usepackage{tikzscale}
\pgfplotsset{compat=newest}
\usetikzlibrary{plotmarks}
\usepackage{rotating}
\usepackage[absolute,overlay]{textpos}
\usepackage{circuitikz}

\usepackage{amsmath}
\usepackage{amssymb}
\usepackage{amsthm}

\newcommand\norm[1]{\left\lVert#1\right\rVert}

\newcommand{\rank}{\operatorname{rank}}
\newcommand{\diag}{\operatorname{diag}}

\DeclareMathOperator*{\minimize}{\text{minimize}}

\newcommand{\R}{\mathbb{R}}

\newcommand{\C}{\mathbb{C}}

\newcommand{\vct}[1]{\boldsymbol{#1}}

\newcommand{\vb}{\vct{b}}

\newcommand{\vf}{\vct{f}}

\newcommand{\vh}{\vct{h}}

\newcommand{\vp}{\vct{p}}
\newcommand{\vq}{\vct{q}}

\newcommand{\vs}{\vct{s}}

\newcommand{\vv}{\vct{v}}
\newcommand{\vw}{\vct{w}}
\newcommand{\vx}{\vct{x}}
\newcommand{\vy}{\vct{y}}

\newcommand{\<}{\langle}
\renewcommand{\>}{\rangle}
\newcommand{\vlambda}{\vct{\lambda}}
\newcommand{\vnu}{\vct{\nu}}

\newcommand{\vzero}{\vct{0}}

\newcommand{\vdelta}{\vct{\delta}}
\newcommand{\mtx}[1]{\boldsymbol{#1}}

\newcommand{\mA}{\mtx{A}}

\newcommand{\mC}{\mtx{C}}
\newcommand{\mD}{\mtx{D}}

\newcommand{\mF}{\mtx{F}}

\newcommand{\mJ}{\mtx{J}}

\newcommand{\mP}{\mtx{P}}
\newcommand{\mQ}{\mtx{Q}}

\newcommand{\mX}{\mtx{X}}

\newcommand{\mnabla}{\mtx{\nabla}}
\newtheorem{lem}{Lemma}

\newtheorem{prop}{Proposition}

\usepackage{graphicx}
\usepackage{amsmath}
\usepackage{array}
\usepackage[caption=false,font=footnotesize]{subfig}
\usepackage{microtype}
\usepackage{dblfloatfix}
\usepackage{cite}
\usepackage{algorithm, algorithmic}
\usepackage{geometry}
\usepackage{authblk}

\begin{document}
\title{Identifiability Conditions for Multi-channel Blind Deconvolution
with Short Filters}

\author[1]{Antoine Paris}
\author[1]{Laurent Jacques}
\affil[1]{ICTEAM/ELEN, UCLouvain, Louvain-la-Neuve, Belgium}
\renewcommand\Affilfont{\itshape\small}
\affil[ ]{Email: \{%
\href{mailto:antoine.paris@uclouvain.be}{\textbf{antoine.paris}},
\href{mailto:laurent.jacques@uclouvain.be}{laurent.jacques}\}@uclouvain.be}

\setcounter{Maxaffil}{0}

\maketitle

\begin{abstract}
	This work considers the multi-channel blind
	deconvolution problem under the assumption that
	the channels are short. 
	First, we investigate the ill-posedness issues inherent to blind
	deconvolution problems and sufficient and necessary conditions
	on the channels that guarantee well-posedness are derived.
	Following previous work on blind deconvolution,
	the problem is then reformulated as a low-rank matrix
	recovery problem and solved by nuclear norm minimization.
	Numerical experiments show the effectiveness of this algorithm
	under a certain generative model for the input signal and the
	channels, both in the noiseless and in the noisy case.
\end{abstract}

\section{Introduction}
The objective of multi-channel blind deconvolution is to recover
unknown $L$-length vectors $\vs$ and $\{\vw_n\}_{n=0}^{N-1}$ from
their circular\footnote{%
In blind deconvolution problems, circular
convolution is generally preferred to linear convolution so that
the convolution theorem for the Discrete Fourier
Transform (DFT) applies. Even though convolutions arising in
practical scenarios are not circular, a linear convolution can
reasonably be approximated by a circular convolution if the support
of the filter is sufficiently short with respect to
the length of the input signal (i.e., if the filter decays
quickly)~\cite{4storiesStrohmer}.
The assumption that the convolution is circular is thus not limiting in
most practical situations.
} convolutions
\begin{equation}
	\vy_n = \vs \circledast \vw_n, \qquad n \in [N],
	\label{eq:simc-bd}
\end{equation}
where $[N]= \{0, \dots, N-1\}$.
This problem has a lot of applications in many fields
of science and engineering. In communications systems for example, blind
deconvolution is referred to as blind equalization~\cite{comm-bdconv}.
In this context, $\vs$ is a message that needs to be
transmitted to a receiver having $N$ antennas
whereas $\vw_n$ is the channel
from the transmitter to the $n$-th receiving antenna.
Traditionally, the transmitter would first
send a \textit{known} pilot message that allows the receiver to estimate
the channels. The receiver would then use these estimates to invert the
effect of the channels on subsequent messages. If the channels are
varying over time (e.g., in mobile communications), this process
needs to be periodically repeated, causing a non-negligible overhead.
Performing blind channel equalization would allow to avoid this~overhead.

Multi-channel blind deconvolution can also be used in noise
imaging applications~\cite{noise-imaging}. One such application
in the field of geophysical imaging is ``seismic-while-drilling'' (SWD).
In SWD, the objective is to perform subsurface imaging on a drilling
site without disrupting drilling operations. One proposed way of
achieving this is to use the vibrations generated by the drill-bit in
the subsurface as the input signal $\vs$. Multiple receivers recording
the seismic traces at several locations would then allow to recover the
``channel impulse responses of the Earth'' $\{\vw_n\}_{n=0}^{N-1}$ using
multi-channel blind deconvolution~\cite{bhar2018fbd}.

Other applications of blind deconvolution include
medical imaging~\cite{medic-bdconv}, astronomical
imaging~\cite{astro-bdconv} and computer vision~\cite{levin11}.

\subsection{Problem setup}
\begin{figure}
    \centering
    \includegraphics[width=0.6\textwidth]{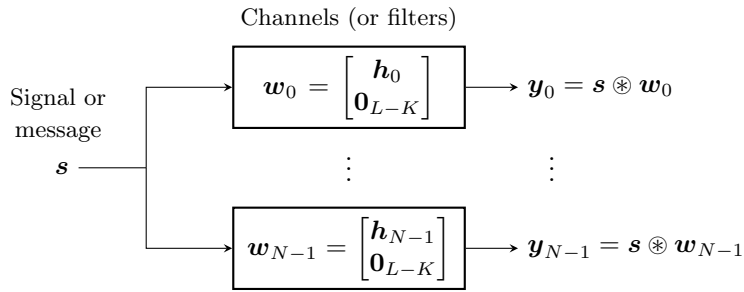}
    \caption{Illustration of the multi-channel blind deconvolution
	    problem under the assumption that the channels are short,
	    i.e., only their $K$ first entries are non-zero.}
    \label{fig:prob-illustration}
\end{figure}

Without any additional assumptions, the multi-channel blind deconvolution
problem formulated above counts more unknowns than observations.
The problem is thus under-determined and ill-posed.
In this work, the channels (also often called filters)
are assumed to live in the following $K$-dimensional subspace
\[
    \vw_n = \mC\vh_n =
    \begin{bmatrix}
        \vh_n \\
	\vzero_{L-K}
    \end{bmatrix},
    \qquad \vh_n \in \R^K, \qquad n \in [N],
\]
where $\mC$ consists
in the first $K$ columns of the $L\times L$ identity matrix. 
In other words, the channels are assumed to be \textit{short}
or supported on $[K]$.
The input signal $\vs$ is simply assumed to be an unknown
\mbox{$L$-length} random vector.
Recovering $\vs$ and $\{\vw_n\}_{n=0}^{N-1}$ is then equivalent
to recovering $\vs$ and $\{\vh_n\}_{n=0}^{N-1}$ and a mandatory condition
for successful recovery reads ${LN \ge L + KN}$, i.e., the information
count must be favorable.
The problem setup is illustrated in~Fig.~\ref{fig:prob-illustration}.

\subsection{Related work and contributions}
Blind deconvolution being an ubiquitous problem in science and
engineering, there exists a vast amount of literature on the topic.
This work was primarily
inspired by~\cite{DBLP:journals/corr/abs-1211-5608,
DBLP:journals/corr/AhmedD16}. In~\cite{DBLP:journals/corr/abs-1211-5608},
Ahmed, Recht and Romberg consider the single-input single-channel blind
deconvolution problem. Both the input signal and the channel are assumed
to live in known subspaces, one of which is random.
In~\cite{DBLP:journals/corr/AhmedD16}, Ahmed and Demanet leverage
multiple diverse inputs -- living in known random subspaces~--
to weaken the assumption on the channel;
the subspace in which the channel
lives is now also to be discovered.
In this work, in a sense ``dual'' to the one studied
in~\cite{DBLP:journals/corr/AhmedD16}, we leverage multiple diverse
and short channels to free the input signal from any assumption.
It is interesting to note that no random subspaces are involved in this work.
This more realistic setting is challenging as it prevents us to use the
classical stability proofs based on, e.g., high dimensional statistics and
measure concentration theory.
The short channel assumption was already used
in~\cite{DBLP:journals/corr/LiLSW16} for the single-input single-channel
case and in~\cite{doi:10.1117/12.2018550,
doi:10.1121/1.4983311} for the multi-channel case.
However,~\cite{doi:10.1121/1.4983311} makes extra-assumptions only valid
for the specific case of underwater acoustic deconvolution.

In~\cite{doi:10.1117/12.2018550,
DBLP:journals/corr/abs-1211-5608,
DBLP:journals/corr/AhmedD16,doi:10.1121/1.4983311}, the problem is
reformulated as a low-rank matrix recovery problem from linear
measurements and solved via nuclear norm minimization using Burer
and Monteiro's algorithm~\cite{doi:10.1137/070697835,Burer2003,Burer2005}.
However,~\cite{doi:10.1117/12.2018550} only shows ``example'' of successful
recovery but lacks a detailed characterization of the performances of the
algorithm, thus giving little insight on the feasibility of the problem.
Robustness to noise is not discussed neither. 
In this work, we also use Burer and Monteiro's algorithm. Yet, and at
the opposite of prior works using this algorithm to solve blind deconvolution
problems, we do not convexify the problem according to the work
in~\cite{Burer2005} and use a heuristic strategy to deal with local minima
instead. This also seems to be the approach followed
in~\cite{doi:10.1121/1.4983311}, although no strategy to deal with local
minima is described.
\textit{The first contribution of this work is to
assess the performances of Burer and Monteiro's algorithm
to solve multi-channel blind deconvolution with short channels}.

Making structural assumptions on
the channels is not sufficient to guarantee that
the problem is well-posed. In~\cite{xu95}, necessary and sufficient
identifiability conditions are derived for multi-channel blind deconvolution
with short filters. \textit{The second contribution of this work is to give a new
interpretation of those conditions by linking them to a localized ill-posedness
through a local analysis -- around the ground truth -- of the $\ell_2$-norm
misfit between the observations and a candidate solution}.

\subsection{Organization of the paper}
Sec.~\ref{sec:theory} first investigates the ill-posedness issues
arising in multi-channel blind deconvolution with short filters.
Necessary and sufficient conditions on the channels that make the
problem well-posed are derived. A class of channels satisfying the
aforementioned conditions almost
surely is also given.
Next, Sec.~\ref{sec:algo} reformulates the multi-channel blind
deconvolution problem as a low-rank matrix recovery problem and
describes Burer and Monteiro's algorithm and the heuristic
strategy used to avoid getting trapped at local minima.
Numerical experiments then assess the performances of this 
algorithm under a generative model for the input signal and the
channels that ensures that the problem is well-posed. 
Both the noiseless and the noisy cases are investigated.
Finally, Sec.~\ref{sec:ccl} concludes this work by discussing the
obtained results, the limits of the short channel assumption, and
some perspectives for future research.

\section{Local ill-posedness and identifiability conditions}
\label{sec:theory}
As discussed in the introduction, assuming $K$-length channels
and having $LN \ge L + KN$ is not sufficient to guarantee that the
problem~\eqref{eq:simc-bd} is well-posed.
Indeed, let $\vf$ be a filter that admits an inverse $\vf^{-1}$ such that
$\vf \circledast \vf^{-1} = \vdelta$. If $\vf$ and $\vw_n$ are such that
$\vw_n' = \vw_n \circledast \vf$ is still supported on $[K]$ for
all $n \in [N]$, then $\{\vw'_n\}_{n=0}^{N-1}$ and
$\vs' = \vs \circledast \vf^{-1}$ also constitutes a valid solution
of the multi-channel blind deconvolution problem. This
ill-posedness will be referred to as \textit{convolutive ambiguity}.
As a particular case, $\{\alpha\vw_n\}_{n=0}^{N-1}$ and $\alpha^{-1}\vs$
also constitutes a valid solution for any scalar $\alpha \neq 0$.
While this \textit{scalar ambiguity}
is acceptable in most applications, convolutive ambiguity is not.
In this section, we derive sufficient and necessary conditions on the
channels that guarantee that the scalar ambiguity is the only possible
ambiguity, making the problem~well-posed.

We start by looking at the observed circular convolutions in the Fourier
domain. Noting $\mF \in \C^{L\times L}$ the unitary DFT
matrix and using the convolution theorem,~\eqref{eq:simc-bd} becomes
\begin{equation}
	\mF\vy_n = \sqrt{L} \mF\vs \odot \mF\mC\vh_n, \qquad n \in [N],
	\label{eq:simc-bd-fourier}
\end{equation}
where $\odot$ is the entry-wise product. 
We then define a bilinear map 
$\mathcal{B}: \C^L \times \R^{KN} \to \C^{LN}$
mapping any input signal in the Fourier domain $\hat{\vp} \in \C^L$ and any set of
$K$-length filters $\{\vq_n\}_{n=0}^{N-1}$ to their corresponding observed
circular convolutions in the Fourier domain,~i.e.,
\[ \mathcal{B}(\hat{\vp}, \{\vq_n\}_{n=0}^{N-1}) =
	\{\sqrt{L}\hat{\vp}[l]\hat{\vq}_n[l]~|~ (l, n) \in [L]\times [N] \}, \]
where $\hat{\vq}_n = \mF\mC\vq_n$. Eq.~\eqref{eq:simc-bd-fourier}
then reads $\hat{\vy} = \mathcal{B}(\hat{\vs}, \{\vh_n\}_{n=0}^{N-1})$
with $\hat{\vy}$ the concatenation of all the observed convolutions in
the Fourier domain and $\hat{\vs} = \mF\vs$.
To ease notations, let $\vq$
be the concatenation of all the filters $\{\vq_n\}_{n=0}^{N-1}$ and
$\vx = [\hat{\vp}^*, \vq^*]^*$ be the concatenation of all the arguments
of $\mathcal{B}$.
The natural way to solve the multi-channel blind deconvolution problem
is then to minimize the $\ell_2$-norm data misfit given by
\[ \textstyle{%
f(\vx) = \frac{1}{2}\norm{\mathcal{B}(\vx) - \hat{\vy}}^2_2.} \]

First, Lemma~\ref{lem:null-space-ill}, proved in
Appendix~\ref{app:proof-null-space-ill}, makes the link between
``local ill-posedness'' and the Hessian of the objective function
$f$ around the ground truth.
\begin{lem}
	\label{lem:null-space-ill}
	Let $\vx_0$ be the ground truth.
	For sufficiently small $\epsilon$,
	$\vx_0 + \epsilon\vv$ also minimizes the objective function
	if $\vv$ belongs to the null space of $\mnabla^2 f(\vx_0)
	\in \C^{(L + KN)\times (L + KN)}$. 
\end{lem}
For ease of notation, the null space of the matrix $\mnabla^2 f(\vx_0)$ is
noted~$\mathcal{N}$. Lemma~\ref{lem:null-space-ill} tells us
that characterizing $\mathcal{N}$ is equivalent to characterizing
the local ill-posedness of the problem around the ground truth.
The following known lemma~\cite{hansen2004nonlinear} is the first
step toward the characterization of $\mathcal{N}$.
\begin{lem}
    \label{lem:hessian}
    Let $\mathcal{A}$ be a differentiable operator such that
    $\mathcal{A}(\vx_0) = \vb$. Let also
    $g(\vx) = \frac{1}{2}\norm{\mathcal{A}(\vx) - \vb}^2_2$.
    The Hessian of $g(\vx)$ at $\vx_0$ is then given by
    \[ \mnabla^2 g(\vx_0) = \mA^*\mA, \]
    where
    $\mA = \frac{\delta \mathcal{A}(\vx)}{\delta \vx}\rvert_{\vx_0}$
    is the linearization of $\mathcal{A}(\vx)$ around
    $\vx_0$ (i.e., the Jacobian of $\mathcal{A}$ at $\vx_0$).
\end{lem}
Using Lemma~\ref{lem:hessian}, we then have $\mnabla^2 f(\vx_0) =
\mJ^*\mJ$ with ${\mJ \in \C^{LN\times(L+KN)}}$ the Jacobian of
$\mathcal{B}$ at $\vx_0$ given by
\[
    \mJ =
    \begin{bmatrix}
        \mD_{\hat{\vw}_0} & \mD_{\hat{\vs}}\mF_K
			  & \mathbf{0} & \cdots & \mathbf{0} \\
        \mD_{\hat{\vw}_1} & \mathbf{0} & \mD_{\hat{\vs}}\mF_K
			  & \cdots & \mathbf{0} \\
        \vdots & \vdots & \vdots & \ddots & \vdots \\
        \mD_{\hat{\vw}_{N-1}} & \mathbf{0} & \mathbf{0} 
			      & \cdots & \mD_{\hat{\vs}}\mF_K
    \end{bmatrix}
\]
where $\mD_{\vx} = \diag(\vx)$ and $\mF_K \in \C^{L\times K}$ is
the restriction of $\mF$ to its $K$ first columns.
The next lemma, proved in
Appendix~\ref{app:proof-null-space-obvious}, gives us a first insight
on the structure of $\mathcal{N}$. 
\begin{lem}
	\label{lem:null-space-obvious}
	The kernel $\mathcal{N}$ is always at least
	one-dimensional and contains the scalar ambiguity inherent to any
	blind deconvolution problem.
\end{lem}
The scalar ambiguity is related to Lemma~\ref{lem:null-space-obvious} as
it imposes the existence of a vector $\vv \in \mathcal{N}$ such that, if
${\vx_0=[\hat{\vs}^*, \vh^*]^*}$ is the ground truth, then, for $\epsilon$
sufficiently small, ${\vx'=[(1+\epsilon)^{-1}\hat{\vs}^*,
(1+\epsilon)\vh^*]^* \approx \vx_0 + \epsilon\vv}$ is also a valid
solution, with $\vv = [-\hat{\vs}^*, \vh^*]^*$ by a first order approximation
(see Appendix~\ref{app:proof-null-space-obvious}).
For the scalar ambiguity to be the only possible ambiguity,
we thus would like $\mathcal{N}$ to be exactly one-dimensional.
From the rank-nullity theorem, a first
necessary condition is to have $LN \ge L + KN - 1$.
The next proposition, proved in Appendix~\ref{app:proof-null-space},
gives two sufficient and necessary conditions on
the filters for $\mathcal{N}$ to be exactly one-dimensional.
\begin{prop}
    \label{prop:null-space}
    Assume that the signal $\vs$ has no zero in the Fourier domain,
    i.e., $\hat{\vs}[l] \neq 0$ for all $l \in [L]$, and
    $LN \ge L + KN - 1$. The kernel $\mathcal{N}$ is one-dimensional
    if and only if
    \begin{enumerate}
        \item there exists an index $n' \in [N]$ such that
	$\vh_{n'}[K-1] \neq 0$,
	\item the polynomials
	$\{\sum_{k=0}^{K-1} \vh_n[k]z^k\}_{n=0}^{N-1}$ 
	are coprime, i.e., they do not share any common root.
    \end{enumerate}
\end{prop}
In particular, Prop.~\ref{prop:null-space} implies that the support
of the channels must be complete on ``both sides'' to avoid ill-posedness
issues.
Indeed, the first condition prevents situations where \mbox{$\vh_n[K-1] = 0$}
for all $n \in [N]$, i.e., situations where the ``end'' of the
support is not filled by any channel. Furthermore, the second condition
prevents, in particular, situations where $\vh_n[0] = 0$ for all $n \in [N]$
(as 0 would then be a shared root), i.e., situations where the ``beginning''
of the support is not filled by any channel. 
As a consequence, the support size $K$
needs to be exactly known to avoid ill-posedness issues.
This limits our short channels model to situations where the
support size of the channels can be \textit{reliably} estimated
beforehand.

As mentioned in the introduction, similar sufficient
and necessary conditions were already derived in~\cite{xu95}.
To our knowledge, this is, however, the first time that those
conditions are linked to $\mathcal{N}$ and to a notion
of ``local ill-posedness''.
In~\cite{xu95}, the authors also derive a condition
involving the input signal~$\vs$. It is not obvious how this condition
might appear from the above reasoning.

The conditions given in Prop.~\ref{prop:null-space} are verified
by specific filters. The next proposition,
proved in Appendix~\ref{app:proof-continous-rv}, determines a
class of random vectors $\{\vh_n\}_{n=0}^{N-1}$ that satisfy the two
conditions of Prop.~\ref{prop:null-space} almost surely.
\begin{prop}
    \label{prop:continous-rv}
    $K$-length vectors $\{\vh_n\}_{n=0}^{N-1}$ whose entries are
    independent continuous random variables whose distribution can be defined by
    density functions satisfy the conditions
    of Prop.~\ref{prop:null-space} almost surely.
\end{prop}

\section{Algorithm and numerical experiments}
\label{sec:algo}
Following the approach described in~\cite{DBLP:journals/corr/AhmedD16},
this section reformulates the multi-channel blind deconvolution problem
and describes the algorithm used to solve the obtained new problem.

The observed convolutions~\eqref{eq:simc-bd} can be written in the
Fourier domain as $\hat{\vy} = \mathcal{A}(\vs\vh^*)$
where $\vh$ is the concatenation of all the unknown filters and
$\mathcal{A}$ is a linear operator.
Multi-channel blind deconvolution can thus be recasted
as a low-rank matrix recovery problem under linear equality
constraints as follows
\[
	\text{find}~\mX
	\quad
	\text{s.t.} 
	\quad
     	\hat{\vy} = \mathcal{A}(\mX)
	\quad\text{and}\quad
     	\rank(\mX) = 1.
\]
This problem is however non-convex and NP-hard.
Fortunately, it can be relaxed as a convex program using the nuclear norm
heuristic~\cite{doi:10.1137/070697835}
\begin{equation}
    \begin{aligned}
	& \underset{\mX}{\minimize} 
        & & \norm{\mX}_*
	& \quad\text{s.t.}\quad
        & & \hat{\vy} = \mathcal{A}(\mX),
    \end{aligned}
    \label{eq:cvx-relax}
\end{equation}
where $\norm{\mX}_*$ is the sum of the singular values of $\mX$.
Because $\mX$ is known to be a rank-1 matrix, we
then write $\mX = \vp\vq^*$ with $\vp$ and $\vq$ the candidate
solution for the input signal and the filters, respectively.
Program~\eqref{eq:cvx-relax} can be shown to be equivalent~to 
\begin{equation}
    \begin{aligned}
        & \underset{\vp, \vq}{\minimize} 
        & & \norm{\vp}^2_2 + \norm{\vq}^2_2 
        & \quad\text{s.t.}\quad
        & & \hat{\vy}= \mathcal{A}(\vp\vq^*),
    \end{aligned}
    \label{eq:low-rank-param}
\end{equation}
except that the latter is non-convex and thus subject to local
minima. In~\cite{Burer2003}, Burer and Monteiro proposed an
algorithm based on the method of multipliers to efficiently find
a local minimum of~\eqref{eq:low-rank-param}.
Starting from an initial guess, this algorithm iteratively
minimizes the following augmented Lagrangian with respect to
$\vp$ and $\vq$~\cite{doi:10.1137/070697835,Burer2003}
\begin{equation*}
    \textstyle{%
    \mathcal{L}(\vp, \vq, \vlambda, \sigma) =
    \frac{1}{2}\left(\norm{\vp}^2_2 +
    \norm{\vq}^2_2\right)
    - \<\vlambda, \mathcal{A}(\vp\vq^*) - \hat{\vy}\>}
    \textstyle{%
    +\frac{\sigma}{2}\norm{\mathcal{A}(\vp\vq^*)
    - \hat{\vy}}^2_2},
\end{equation*}
where $\vlambda \in \C^{LN}$ is the vector of Lagrange multipliers and
$\sigma$ is a positive penalty parameter. A way of updating $\vlambda$
and $\sigma$ at each iteration is also proposed in~\cite{Burer2003}.
The algorithm stops when
$\norm{\mathcal{A}(\vp\vq^*) - \hat{\vy}}^2_2$ is below a
given tolerance. In this work, the
minimization is performed using L-BFGS with the \textsc{MATLAB} solver
\textit{minFunc}~\cite{minFunc}.

To avoid getting trapped at local minima and find a global minimum
instead, the following heuristic strategy was implemented. When
$\norm{\mathcal{A}(\vp\vq^*) - \hat{\vy}}^2_2$ stopped decreasing for
a given number of iterations and is largely above the tolerance,
the algorithm decrees to be trapped at a local minimum.
At this point, the algorithm simply starts again from
another initial guess.
Following the work of Burer and
Monteiro~\cite{Burer2005},~\eqref{eq:low-rank-param} can also be
convexified by writing $\mX = \mP\mQ^*$ with $\mP \in \R^{L\times r}$,
$\mQ \in \R^{KN\times r}$ and $r > 1$. In this case, the algorithm stops
when either $\mP$ or $\mQ$ is rank deficient and the recovered vectors
$\vp$ and $\vq$ are given by the leading singular vectors of $\mP$ and
$\mQ$, respectively.
This is for example the approach used
in~\cite{DBLP:journals/corr/abs-1211-5608} for the single-input
single-channel case.
However, we empirically observed
better performances with the first approach.
Interestingly, this is also the
approach followed in~\cite{doi:10.1121/1.4983311} to solve multi-channel
blind deconvolution for underwater acoustic, although 
no strategy to handle the local minima is described.

In the absence of a better strategy, $\vp$
and $\vq$ are initialized as independent standard random Gaussian vectors.
The deterministic nature of the
linear operator $\mathcal{A}$ does not allow us to use a smarter
initialization scheme such as, for example, the spectral method
proposed~in~\cite{DBLP:journals/corr/LiLSW16}.

A sketch of the complete algorithm is given in Algorithm~\ref{alg:bd}.

\begin{algorithm}
	\caption{Burer and Monteiro's algorithm with heuristic
	local minima detection}
	\label{alg:bd}
 	\begin{algorithmic}[1]
 	\renewcommand{\algorithmicrequire}{\textbf{Input:}}
 	\renewcommand{\algorithmicensure}{\textbf{Output:}}
	\REQUIRE $\hat{\vy} = \mathcal{A}(\vs\vh^*)$
	\ENSURE $(\vp, \vq)$, estimate for $(\vs, \vh)$
 	%\\ \textit{Initialization}:
  	\STATE $\vp, \vq \leftarrow$ independent standard random Gaussian
	vectors
	\STATE $\vlambda = \vzero$, $\sigma = \sigma_0$
 	%\\ \textit{Optimization}:
	\WHILE {$\norm{\mathcal{A}(\vp\vq^*) - \hat{\vy}}^2_2 > \delta$}
	\STATE Minimize $\mathcal{L}(\vp, \vq, \vlambda, \sigma)$ w.r.t.
	$\vp$ and $\vq$ with L-BFGS
	\STATE Update $\vlambda$ and $\sigma$ according to~\cite{Burer2003}
  	\IF {trapped at a local minima}
  	\STATE Go back to step 1
  	\ENDIF
  	\ENDWHILE
	\RETURN $(\vp, \vq)$ 
 	\end{algorithmic} 
\end{algorithm}

\subsection{Numerical experiments}
\label{sec:num-exp}
In the following numerical experiments, both the input signal~$\vs$
and the channels~$\{\vh_n\}_{n=0}^{N-1}$ are independent random Gaussian
vectors width i.i.d. entries.
In this way, the identifiability conditions of
Prop.~\ref{prop:null-space} are satisfied almost surely. To comply with
the existing literature on blind deconvolution using low-rank matrix
recovery, the metric used to assess the quality of the recovery is the
relative outer product error, i.e,
\[ \norm{\mX_0 - \vp\vq^*}_F\big/\norm{\mX_0}_F, \]
where $\mX_0 = \vs\vh^*$ is the ground truth, $\vp$ is the recovered
input signal and $\vq$ is the concatenation of the recovered filters.

\subsubsection{Noiseless case}
\label{sec:phase-transitions}
\begin{figure*}
	\centering
	\subfloat[
		Empirical probability of successful recovery in the
		noiseless case. White means a 100\% probability while
		black means a 0\% probability.	
	]{\includegraphics[width=0.42\textwidth]{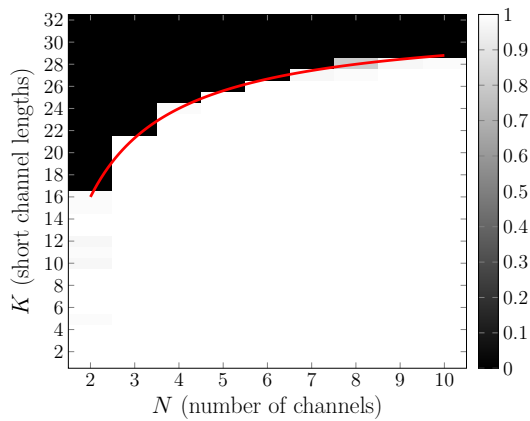}%
	\label{fig:KN32}}
	\hfill
	\subfloat[
		Average number of attempts before recovery in the
		noiseless case. Darker means that more attempts were
		needed to succeed in average..
	]{\includegraphics[width=0.42\textwidth]{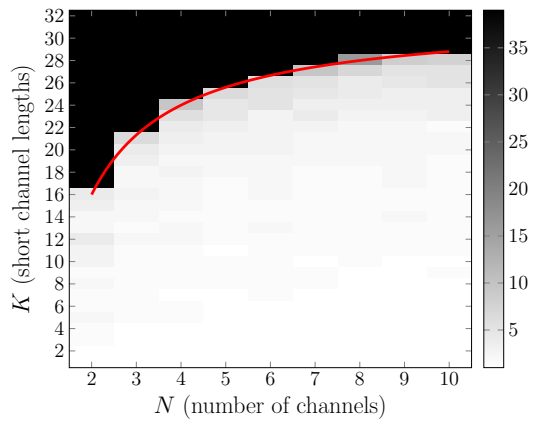}%
	\label{fig:KN32-num-att}}
    	\caption{Results of the numerical experiments with independent standard
		Gaussian vectors $\vs$ and $\{\vh_n\}_{n=0}^{N-1}$ with i.i.d.
		entries.}
	\label{fig:num-exp}
\end{figure*}

We first assess the performances of the proposed algorithm in the
noiseless case.
Figures~\ref{fig:KN32} and~\ref{fig:KN32-num-att} depict the empirical
probability of successful recovery and the average number of attempts
before successful recovery (i.e., the number of times the algorithm
has to restart because it got stuck at a local minimum)
when $L = 32$, $N$ is ranging from 2 to 10 and $K$ is ranging from 1 to 32.
The recovery is considered to be successful if the relative outer
product error is less than\footnote{This threshold has been chosen
to comply with the existing literature. Most of the time, however, the
actual relative error is orders of magnitude smaller.} 2\%.
For each pair of parameters $N$ and $K$, the value displayed
in Figs.~\ref{fig:KN32} and~\ref{fig:KN32-num-att} is the result of an
average over 100 deconvolutions, each time involving new random instances
of $\vs$ and $\{\vh_n\}_{n=0}^{N-1}$.
On both figures, a red curve indicates the information-theoretic
limit. Above this curve, the problem is under-determined and successful
recovery is thus not possible. Below this curve, however,
Fig.~\ref{fig:KN32} shows that the recovery is successful with
probability 1 almost everywhere.
This result is particularly interesting when compared to ``usual''
results in blind deconvolution where the minimal number of observations
is larger than the number of unknowns (usually within log factors).
Next, Fig.~\ref{fig:KN32-num-att} shows that the average number
of attempts before successful recovery is usually small and decreases
when the oversampling factor increases.
The decision to not convexify~\eqref{eq:low-rank-param}
along with the heuristic strategy implemented to avoid getting
trapped at local minima thus seem reasonable.

\subsubsection{Robustness to noise}
\label{sec:noise-robustness}
\begin{figure}
    \centering
	\includegraphics[width=0.5\textwidth]{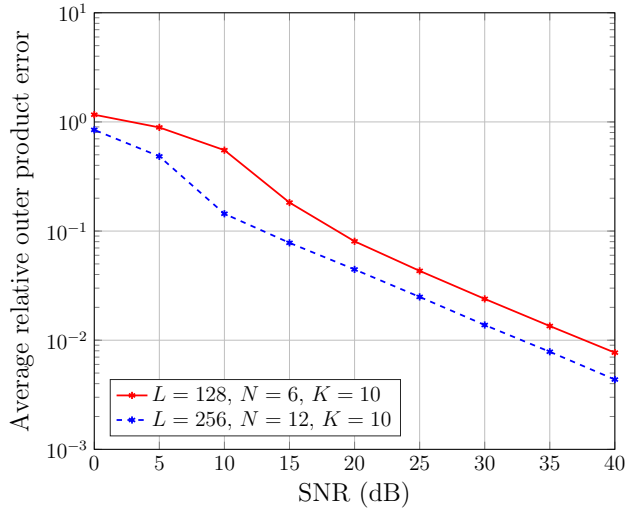}%
    \caption{Average relative outer product error against the SNR. 
    Except at low SNR, the relative error scales linearly with the noise
    level as desired.}
	\label{fig:noise-robustness}
\end{figure}

Each observed convolution $\vy_n$ is now corrupted by an additive
noise $\vv_n$ given by
\[ \vv_n = \sigma \cdot \norm{\vy_n}_2 \cdot
\vnu_n \big/\norm{\vnu_n}_2, \]
where $\vnu_n \in \R^L$ is a standard Gaussian random vector.
The SNR is then simply given by $\text{SNR}_n = -20\log_{10}\sigma$
and is assumed to be the same for each observed convolution.
Fig.~\ref{fig:noise-robustness} depicts the average
relative outer product error against the SNR. Each data point is the
result of an average over 2800 deconvolutions. For each deconvolution,
new random instances of $\vs$, $\{\vh_n\}_{n=0}^{N-1}$ and the noise
were used. Except at low SNR (i.e., less than 10dB), the relative error
scales linearly with the noise level in a log-log scale as desired.
In addition, Fig.~\ref{fig:noise-robustness} shows that
increasing the oversampling factor increases the robustness to noise.

\section{Conclusion}
\label{sec:ccl}
Section~\ref{sec:theory} revisited known identifiability conditions
for the channels and linked those conditions with a notion of
``local ill-posedness'' around the ground truth. It is not yet
clear, however, how identifiability conditions involving the input signal 
could be derived with the same approach. This section also
highlighted a limitation of the short channels model used in
this work; the support size $K$ needs to be \textit{exactly} known
for the problem to be well-posed. In most applications, however,
the support size is not known a priori but rather needs to be discovered.
In such applications, other ways of encoding the short channels assumption
should be employed (see for example~\cite{bhar2018fbd}).
What happens when the conditions of Prop.~\ref{prop:null-space} are
``almost not satisfied'' (e.g., when the support is ``almost not filled''
or when the polynomials $\{\sum_{k=0}^{K-1} \vh_n[k]z^k\}_{n=0}^{N-1}$
``almost share a root'') is another interesting question.
Preliminary numerical experiments suggest that it
degrades well-posedness. However, how to quantitatively measure
this degradation remains an open question.

Next, Sec.~\ref{sec:num-exp} demonstrated the effectiveness of the
proposed algorithm to solve multi-channel blind deconvolution under
a generative model for the channels that ensures that the problem is
well-posed. It would also be interesting to study the
impact of channels similarities on the performances of the algorithm.
To this effect, a measure of channels similarities and
a way of controlling it must first be devised.
The condition number of the $K\times N$ matrix whose
columns are the channels might be a good candidate
to measure channels similarities, as suggested by unreported experiments.

\section*{Acknowledgment}
The authors would like to thank Laurent Demanet, Ali Ahmed and
Pawan Bharadwaj from MIT and Luc Vandendorpe from UCLouvain for
interesting discussions. Laurent Jacques is funded by the F.R.S.-FNRS.

\appendix

\section{Proof of Lemma~\ref{lem:null-space-ill}}
\label{app:proof-null-space-ill}
We write the Taylor series of the objective function around the ground
truth $\vx_0$
\[\textstyle{%
    f(\vx_0 + \epsilon\vv) = f(\vx_0) + \epsilon\mnabla f(\vx_0)^T\vv
    + \frac{\epsilon^2}{2}\vv^*\mnabla^2 f(\vx_0)\vv + \cdots}.
\]
Assuming $\epsilon$ is small enough for the higher order terms to be
negligible and because $f(\vx_0) = 0$ and $\mnabla f(\vx_0) = 0$,
points $\vx_0 + \epsilon\vv$ such that
\begin{equation}
    f(\vx_0 + \epsilon\vv) = 0 \quad \text{or} \quad \vv^*\mnabla^2
    f(\vx_0)\vv = 0
    \label{eq:hess-null-space}
\end{equation}
also minimize the objective function. The set of solutions of this last
equation is given by the null space of $\mnabla^2 f(\vx_0)$.

\section{Proof of Lemma~\ref{lem:null-space-obvious}}
\label{app:proof-null-space-obvious}
The null space of $\mnabla^2 f(\vx_0)$ can be computed by
solving~\eqref{eq:hess-null-space}. Letting
$\vv = [\hat{\vp}^*, \vq^*_0, \dots, \vq_{N-1}^*]^*$, this can
be developed as
\begin{equation}
	\textstyle{%
    \sum_{n=0}^{N-1} \norm{\mD_{\hat{\vw}_n}\hat{\vp} +
    \mD_{\hat{\vs}}\mF_K\vq_n}^2_2 = 0}
    \label{eq:null-space}
\end{equation}
or, equivalently $\mD_{\hat{\vw}_n}\hat{\vp} = -\mD_{\hat{\vs}}\mF_K\vq_n$
for all $n \in [N]$. The obvious solution to this system of equations
is given by ${\hat{\vp} = -\alpha\hat{\vs}}$ and $\vq_n = \alpha\vh_n$ for all
$n \in [N]$ and for any scalar $\alpha$. This gives us a basis vector
of the null space of $\mnabla^2 f(\vx_0)$
\[ \vv = [-\hat{\vs}^*, \vh_0^*, \dots, \vh_{N-1}^*]^*. \]
We now want to prove that the space generated by this vector contains
the scalar ambiguity inherent to any blind deconvolution problem.
Let's consider the ground truth $\vx_0$ and another minimizer of the
objective function resulting from the scalar ambiguity,
$\vx' = [\alpha^{-1}\hat{\vs}^*, \alpha\vh_0^*, \dots,
\alpha\vh_{N-1}^*]^*$ for any scalar $\alpha \neq 0$.
The vector joining $\vx'$ and $\vx_0$ is given by
\[ \vx_0 - \vx' = [\hat{\vs}^*(1 - \alpha^{-1}), \vh_0^*(1-\alpha),
	\dots, \vh_{N-1}^*(1-\alpha)]^*. \]
For $\alpha = 1 + \epsilon$ and $\epsilon$ close to 0, this reduces
to\footnote{Using a first order Taylor series around 0,
$\frac{\epsilon}{1+\epsilon} \approx \epsilon$.} 
\[ \vx_0 - \vx' = [\epsilon\hat{\vs}^*, -\epsilon\vh_0^*,
\dots, -\epsilon\vh_{N-1}^*]^* = -\epsilon\vv \]
and $\vx' = \vx_0 + \epsilon\vv$.

\section{Proof of Prop.~\ref{prop:null-space}}
\label{app:proof-null-space}
The objective is to prove that the obvious solution
to~\eqref{eq:null-space} is the only one.
For other solutions to exist, we would need
$\hat{\vp} = -\mD_{\hat{\vw}_n}^{-1}\mD_{\hat{\vs}}\mF_K{\vq}_n$
for all $n \in [N]$ which then implies
\begin{equation*}
    \mD_{\hat{\vw}_n}^{-1}\mD_{\hat{\vs}}\mF_K\vq_n 
    = \mD_{\hat{\vw}_m}^{-1}\mD_{\hat{\vs}}\mF_K\vq_m,
    \qquad (n, m) \in [N]\times[N].
\end{equation*}
With the assumption that $\hat{\vs}[l] \neq 0$ for $l \in [L]$,
this gives
\begin{equation*}
    \mD_{\hat{\vw}_m}\mF_K\vq_n 
    = \mD_{\hat{\vw}_n}\mF_K\vq_m, \qquad (n, m) \in [N]\times[N] 
\end{equation*}
which in the time domain becomes
\begin{equation}
    \vw_m \circledast \mC\vq_n = \vw_n \circledast \mC\vq_m,
    \qquad (n, m) \in [N]\times[N].
    \label{eq:null-space-conv}
\end{equation}
We now have to prove that $\vq_n = \alpha\vh_n$ for
all $n \in [N]$ is the only solution to this equation if and only
if conditions 1 and 2 are true.
Taking the $\mathcal{Z}$-transform on both sides
of~\eqref{eq:null-space-conv} and using the convolution theorem leads to
\[ \mathcal{Z}(\vw_m)\mathcal{Z}(\mC\vq_n)
= \mathcal{Z}(\vw_n)\mathcal{Z}(\mC\vq_m) \]
for all pairs $(n, m) \in [N]\times[N]$.
After the inconsequential change of variable $z^{-1} \to z$,
and noting $P_{\vx}(z) = \sum_{k=0}^{K-1} \vx[k]z^k$, this becomes
\begin{equation}
    P_{\vh_m}(z)P_{\vq_n}(z) = P_{\vh_n}(z)P_{\vq_m}(z),
    \quad (n, m) \in [N]\times[N]. 
    \label{eq:poly-eq}
\end{equation}
The objective has now been transformed into proving
that~\eqref{eq:poly-eq}
only admits $P_{\vq_n}(z) = \alpha P_{\vh_n}(z)$ for all $n \in [N]$
as a solution if and only if conditions 1 and 2 are true.
    
%\paragraph{Forward direction}
Let us first prove the forward direction by contraposition.
We want to prove that if the negation of either the first or
the second condition is true, then there
exists other solutions to~\eqref{eq:null-space} and $\mathcal{N}$
is thus more than one dimensional.

\begin{enumerate}
    	\item We first assume that the negation of the first condition
	is true: there does not exist any index $n \in [N]$ such
	that $\vh_n[K-1] \neq 0$. Equivalently, there
    	does not exist any index $n \in [N]$ such that
	$P_{\vh_n}(z)$ is of degree $> K-2$. In this case, 
	\[ P_{\vq_n}(z) = \alpha(z-\beta)P_{\vh_n}(z), 
	\qquad n \in [N] \]
	also constitutes a solution to~\eqref{eq:poly-eq} for any
	$\beta \in \C$ and the obvious solution is thus not the only one.
    	\item We now suppose that the negation of the second condition
	is true: there is a root $\beta \in \C$ shared by all the
	polynomials $P_{\vh_n}(z)$. Then
    	\[ P_{\vq_n}(z) = \alpha \frac{P_{\vh_n}(z)}{z-\beta}(z-\gamma),
	\qquad n \in [N] \]
    	also constitutes a solution to~\eqref{eq:poly-eq} for any
	$\gamma \in \C$. For $\gamma \neq \beta$, this solution is
	different from the obvious one and the obvious solution is thus
	not the only one.
\end{enumerate}

%\paragraph{Backward direction}
Next, concerning the backward direction, we want to prove that if the
two conditions are true, then the only solution to~\eqref{eq:null-space}
is the obvious one and $\mathcal{N}$ is one-dimensional.
A necessary condition for~\eqref{eq:poly-eq} to be
satisfied is that $P_{\vh_m}(z)P_{\vq_n}(z)$ and
$P_{\vh_n}(z)P_{\vq_m}(z)$ have the same roots for
all $(n, m) \in [N]\times[N]$. 

\begin{enumerate}
    	\item We first assume that the second condition is satisfied:
	the polynomials $\{P_{\vh_n}(z)\}_{n=0}^{N-1}$ do not share any
	common root. In this case, the necessary condition
	for~\eqref{eq:poly-eq} to be satisfied directly translates to
	$P_{\vq_n}(z)$ having at least all the roots of $P_{\vh_n}(z)$,
	$\forall n \in [N]$. This is however not sufficient to conclude
	that $P_{\vq_n}(z) = \alpha P_{\vh_n}(z)$ for all $n \in [N]$ as
	we could for example add a root $\beta \in \C$ in every
	$P_{\vq_n}(z)$ if the first condition is not true (as was done
	in the preceding paragraph).
    	\item Next, we suppose that the first condition is also
	satisfied: there exists an index $n' \in [N]$ such that
	$\vh_{n'}[K-1] \neq 0$. Equivalently, there exists
    	an index $n' \in [N]$ such that $P_{\vh_{n'}}(z)$ is of degree
	$K-1$. In this case, it is no longer possible to add a root
	$\beta \in \C$ in every $P_{\vq_n}(z)$ as it would result in
	$P_{\vq_{n'}}(z)$ being of degree $K$, which is not possible.
\end{enumerate} 
We can conclude that, if the two conditions are true,
$P_{\vq_n}(z)$ and $P_{\vh_n}(z)$ must have \textit{exactly} the same
roots for all $n \in [N]$. Because polynomials having exactly the same
roots are equal up to a scalar factor, we can conclude that
$P_{\vq_n}(z) = \alpha P_{\vh_n}(z)$ for all $n \in [N]$
is the only solution to~\eqref{eq:null-space} and that $\mathcal{N}$
is one dimensional and only contains the scalar
ambiguity inherent to blind deconvolution.

\section{Proof of Prop.~\ref{prop:continous-rv}}
\label{app:proof-continous-rv}
Let us first prove that the first condition of Prop.~\ref{prop:null-space}
is satisfied with probability one.
For all $n \in [N]$, $\vh_n[K-1]$ is a continuous random variable whose
distribution can be defined by a density function. As such, one can
write
\[ \Pr(\vh_n[K-1] = 0) \overset{a.s.}{=} 0, \qquad n \in [N]\]
and the first condition is satisfied almost surely.
For the second condition, we work with the set of polynomials
$\{P_{\vh_n}(z)\}_{n=0}^{N-1}$.
A proof that the probability for two such polynomials to have any common
root is zero is given in~\cite[p. 442]{PLMS:PLMS0439}. This proof can be
extended to any number of polynomials and the second condition is thus
also satisfied with probability one.

\bibliographystyle{IEEEtran}
\bibliography{main.bib}

\begin{thebibliography}{10}
\providecommand{\url}[1]{#1}
\csname url@samestyle\endcsname
\providecommand{\newblock}{\relax}
\providecommand{\bibinfo}[2]{#2}
\providecommand{\BIBentrySTDinterwordspacing}{\spaceskip=0pt\relax}
\providecommand{\BIBentryALTinterwordstretchfactor}{4}
\providecommand{\BIBentryALTinterwordspacing}{\spaceskip=\fontdimen2\font plus
\BIBentryALTinterwordstretchfactor\fontdimen3\font minus
  \fontdimen4\font\relax}
\providecommand{\BIBforeignlanguage}[2]{{%
\expandafter\ifx\csname l@#1\endcsname\relax
\typeout{** WARNING: IEEEtran.bst: No hyphenation pattern has been}%
\typeout{** loaded for the language `#1'. Using the pattern for}%
\typeout{** the default language instead.}%
\else
\language=\csname l@#1\endcsname
\fi
#2}}
\providecommand{\BIBdecl}{\relax}
\BIBdecl

\bibitem{4storiesStrohmer}
T.~Strohmer, ``{Four short stories about Toeplitz matrix calculations},''
  \emph{Linear Algebra and its Applications}, vol. 343-344, pp. 321--344, 2002.

\bibitem{comm-bdconv}
X.~Wang and H.~V. Poor, ``Blind equalization and multiuser detection in
  dispersive {CDMA} channels,'' \emph{IEEE Transactions on Communications},
  vol.~46, no.~1, pp. 91--103, 1998.

\bibitem{noise-imaging}
J.~Garnier and G.~Papanicolaou, \emph{Passive Imaging with Ambient Noise},
  1st~ed.\hskip 1em plus 0.5em minus 0.4em\relax New York, NY, USA: Cambridge
  University Press, 2016.

\bibitem{bhar2018fbd}
P.~Bharadwaj, L.~Demanet, and A.~Fournier, ``Focused blind deconvolution of
  interferometric {Green's} functions,'' in \emph{SEG Technical Program
  Expanded Abstracts 2018}, 2018, pp. 4085--4090.

\bibitem{medic-bdconv}
O.~Michailovich and A.~Tannenbaum, ``Blind deconvolution of medical ultrasound
  images: A parametric inverse filtering approach,'' \emph{IEEE Transactions on
  Image Processing}, vol.~16, no.~12, pp. 3005--3019, 2007.

\bibitem{astro-bdconv}
S.~M. {Jefferies} and J.~C. {Christou}, ``{Restoration of Astronomical Images
  by Iterative Blind Deconvolution},'' \emph{Astrophysical Journal}, vol. 415,
  pp. 862--874, 1993.

\bibitem{levin11}
A.~Levin, Y.~Weiss, F.~Durand, and W.~T. Freeman, ``Understanding blind
  deconvolution algorithms,'' \emph{IEEE Transactions on Pattern Analysis and
  Machine Intelligence}, vol.~33, no.~12, pp. 2354--2367, 2011.

\bibitem{DBLP:journals/corr/abs-1211-5608}
A.~Ahmed, B.~Recht, and J.~Romberg, ``Blind deconvolution using convex
  programming,'' \emph{IEEE Transactions on Information Theory}, vol.~60,
  no.~3, pp. 1711--1732, 2014.

\bibitem{DBLP:journals/corr/AhmedD16}
A.~Ahmed and L.~Demanet, ``Leveraging diversity and sparsity in blind
  deconvolution,'' \emph{IEEE Transactions on Information Theory}, vol.~64,
  no.~6, pp. 3975--4000, 2018.

\bibitem{DBLP:journals/corr/LiLSW16}
X.~Li, S.~Ling, T.~Strohmer, and K.~Wei, ``Rapid, robust, and reliable blind
  deconvolution via nonconvex optimization,'' \emph{Applied and Computational
  Harmonic Analysis}, 2018.

\bibitem{doi:10.1117/12.2018550}
J.~Romberg, N.~Tian, and K.~Sabra, ``Multichannel blind deconvolution using low
  rank recovery,'' \emph{Proceedings of SPIE}, vol. 8750, pp. 8750 -- 8750 --
  6, 2013.

\bibitem{doi:10.1121/1.4983311}
N.~Tian, S.-H. Byun, K.~Sabra, and J.~Romberg, ``Multichannel myopic
  deconvolution in underwater acoustic channels via low-rank recovery,''
  \emph{The Journal of the Acoustical Society of America}, vol. 141, no.~5, pp.
  3337--3348, 2017.

\bibitem{doi:10.1137/070697835}
B.~Recht, M.~Fazel, and P.~A. Parrilo, ``Guaranteed minimum-rank solutions of
  linear matrix equations via nuclear norm minimization,'' \emph{SIAM Review},
  vol.~52, no.~3, pp. 471--501, 2010.

\bibitem{Burer2003}
S.~Burer and R.~D. Monteiro, ``A nonlinear programming algorithm for solving
  semidefinite programs via low-rank factorization,'' \emph{Mathematical
  Programming}, vol.~95, no.~2, pp. 329--357, 2003.

\bibitem{Burer2005}
------, ``Local minima and convergence in low-rank semidefinite programming,''
  \emph{Mathematical Programming}, vol. 103, no.~3, pp. 427--444, 2005.

\bibitem{xu95}
G.~Xu, H.~Liu, L.~Tong, and T.~Kailath, ``A least-squares approach to blind
  channel identification,'' \emph{IEEE Transactions on Signal Processing},
  vol.~43, no.~12, pp. 2982--2993, 1995.

\bibitem{hansen2004nonlinear}
P.~Hansen, V.~Pereyra, and G.~Scherer, \emph{Least Squares Data Fitting with
  Applications}.\hskip 1em plus 0.5em minus 0.4em\relax John Hopkins University
  Press, 2012.

\bibitem{minFunc}
M.~Schmidt, ``{minFunc: unconstrained differentiable multivariate optimization
  in Matlab},'' \url{http://www.cs.ubc.ca/~schmidtm/Software/minFunc.html},
  2005.

\bibitem{PLMS:PLMS0439}
J.~E.~A. Dunnage, ``The number of real zeros of a class of random algebraic
  polynomials,'' \emph{Proceedings of the London Mathematical Society}, vol.
  s3-18, no.~3, pp. 439--460, 1968.

\end{thebibliography}

\end{document}